\documentclass{easychair}

\usepackage{doc}
\usepackage{oldlfont,amssymb,epsf,stmaryrd,epsfig,color}
\usepackage[utf8]{inputenc}
\usepackage{subfigure}
\usepackage{caption}
\usepackage{tikz}
\usetikzlibrary{arrows,shapes,decorations.pathmorphing,backgrounds,positioning,fit, calc}
\definecolor{blue-violet}{rgb}{0.54, 0.17, 0.89}

\title{{\sc Logipedia}: a multi-system encyclopedia of formal proofs}
\titlerunning{{\sc Logipedia}}

\author{Gilles Dowek \and Fran\c{c}ois Thir\'e}
\authorrunning{Dowek and Thir\'{e}}

\institute{Inria and \'Ecole normale sup\'erieure de Paris-Saclay\\ LSV,
61, avenue du Pr\'esident Wilson
94235 Cachan Cedex, France\\
\email{gilles.dowek@ens-paris-saclay.fr} \\
\email{francois.thire@lsv.fr}}

\begin{document}
\maketitle

Libraries of formal proofs are an important part of our mathematical
heritage, but their usability and sustainability is poor.  Indeed,
each library is specific to a proof system, sometimes even to some
version of this system. Thus, a library developed in one system
cannot, in general, be used in another and when the system is no more
maintained, the library may be lost.  This impossibility of using a
proof developed in one system in another has been noted for long and a
remediation has been proposed: as we have empirical evidence that most
of the formal proofs developed in one of these systems can also be
developed in another, we can develop a standard language, in which
these proofs can be translated, and then used in all systems
supporting this standard.

{\sc Logipedia} ({\tt http://logipedia.science}) is an attempt to
build such a multi-system on-line encyclopedia of formal proofs
expressed in such as standard language. It is based on two main ideas:
the use of a logical framework and of reverse mathematics.

\paragraph{Logical frameworks}

Different proof systems, such as {\sc Coq}, {\sc Matita}, {\sc HOL
Light}, {\sc Isabelle/HOL}, {\sc PVS}... implement variants of
different {\em logical formalisms}: the Calculus of constructions,
Simple type theory, Simple type theory with predicate
subtyping... After several decades of research, we understand the
relationship between these formalisms much better. But, to build an
encyclopedia of formal proofs, we have been one step further and
expressed all these formalisms as theories in a common {\em logical
  framework}.

The idea of using a logical framework, such as predicate logic, to
express theories, such as geometry and set theory, goes back to
Hilbert and Ackermann,
but several other
logical frameworks such as {\em $\lambda$-Prolog},
{\em Isabelle},
{\em the $\lambda \Pi$-calculus}
{\em deduction modulo theory},
and {\em the $\lambda \Pi$-calculus modulo theory}
\cite{DBLP:conf/tlca/CousineauD07,NPS}, have been proposed to solve
some issues of predicate logic. We have used the {\em the
  $\lambda \Pi$-calculus modulo theory}, implemented in the system
{\sc Dedukti} \cite{10auteurs}. With respect to predicate logic this
logical framework allows
\begin{itemize}
\item binders (like {\em
    $\lambda$-Prolog}, {\em Isabelle}, and {\em the
    $\lambda \Pi$-calculus})

\item proofs as $\lambda$-terms (like {\em the
  $\lambda \Pi$-calculus})

\item an arbitrary definitional / computational equality
(like {\em deduction modulo theory})

\item arbitrary connectives and quantifiers, in particular a mix of
  constructive classical ones (like ecumenical logics
  \cite{PereiraRodriguez}).
\end{itemize}

It permits to express Simple type theory with 8 symbol declarations and 3
rewrite rules and the Calculus of constructions with 9 symbol declarations
and 4 rewrite rules \cite{DBLP:journals/corr/abs-1712-01485}.

Expressing the theory implemented in {\sc HOL Light} and
{\sc Matita}
theories D[HOLL] and D[Mat] in {\sc Dedukti} allows to translate proofs
developed in these systems to these theories and back.


\paragraph{Reverse mathematics}

\begin{figure}
  \centering
  \begin{minipage}{.5\textwidth}
    \scalebox{0.65}{
    \begin{tikzpicture}
      [node distance=3cm,
      on grid,
      checker/.style = {shape=ellipse, draw=blue-violet!100, fill=blue-violet!25, align=center,
        minimum height=1.2cm, minimum width=2cm},
      dedukti node/.style = {shape=ellipse, draw=blue-violet!90, fill=blue-violet!50, label={south:Dedukti}},
      post/.style={->, >=stealth', semithick}]

      \node [checker, fill=blue-violet!75, label={center:D[HOLL]}] (DOT) {};
      \node [checker, left=of DOT, fill=blue-violet!75, label={center:D[Mat]}] (DMatita) {};
      \begin{scope}[on background layer]
        \node [dedukti node, inner sep=7pt, fit={(DMatita) (DOT)}] (Dedukti) {};
      \end{scope}
      \node [draw=none] (Coq) [above=of Dedukti] {};
      \node [checker, label={north:HOL-Light}] (OT) [right=of Coq, yshift=-1cm] {};
      \node [checker, label={north:Matita}] (Matita) [left=of Coq, yshift=-1cm] {};
      \draw[orange, very thick, <->] (DOT) -- (OT);
      \draw[orange, very thick, <->] (DMatita) -- (Matita);
    \end{tikzpicture}
}
    \caption{Logical Framework}
  \end{minipage}
  \begin{minipage}{.4\textwidth}
    \scalebox{0.65}{
    \begin{tikzpicture}
      [node distance=3cm,
      on grid,
      checker/.style = {shape=ellipse, draw=blue-violet!100, fill=blue-violet!25, align=center,
        minimum height=1.2cm, minimum width=2cm},
      dedukti node/.style = {shape=ellipse, draw=blue-violet!90, fill=blue-violet!50, label={south:Dedukti}},
      post/.style={->, >=stealth', semithick}]

      \node [checker, fill=blue-violet!75, label={center:D[HOLL]}] (DOT) {};
      \node [checker, left=of DOT, fill=blue-violet!75, label={center:D[Mat]}] (DMatita) {};
      \begin{scope}[on background layer]
        \node [dedukti node, inner sep=7pt, fit={(DMatita) (DOT)}] (Dedukti) {};
      \end{scope}
      \node [draw=none] (Coq) [above=of Dedukti] {};
      \node [checker, label={north:HOL-Light}] (OT) [right=of Coq, yshift=-1cm] {};
      \node [checker, label={north:Matita}] (Matita) [left=of Coq, yshift=-1cm] {};
      \draw[circle,fill=violet!50] (Matita.center) circle (0.25) {};
      \draw[post, very thick, ->, orange] (Matita.center)
      .. controls ([yshift=-0cm,xshift=-3cm]Dedukti)
      .. (DMatita);
      \draw[post, very thick, <-, orange] (OT)
      .. controls ([yshift=-0cm,xshift=3cm]Dedukti)
      .. (DOT);

      \draw[very thick, orange, dashed, ->] (DMatita) .. controls ([xshift=-0.5cm,yshift=0.5cm]Dedukti) and ([xshift=0.5cm,yshift=0.5cm]Dedukti) .. (DOT);
    \end{tikzpicture}
    }
    \caption{Reverse mathematics}
  \end{minipage}
\end{figure}

Expressing various formalisms as theories in the same logical
framework permits to compare them. For instance comparing the
expression of Simple type theory and of the Calculus of constructions
in {\sc Dedukti} \cite{DBLP:journals/corr/abs-1712-01485},
we notice that there are only three
differences the symbol {\em arrow} that permits to build functional
types is dependent in the Calculus of constructions, and the same
holds for the implication $\Rightarrow$. Then the Calculus of
constructions has one extra symbol $\pi$ to build types for functions
mapping proofs to terms and Simple type theory does not.
Thus all proofs expressed in D[HOLL] can be translated to D[Mat].

Conversely, we can identify a subset of the proofs of D[Mat], that
do not use the dependency of the symbols {\em arrow} and $\Rightarrow$
and do not use the symbol $\pi$, and can be translated to
D[HOLL]. Composing these translation we obtain a partial function
mapping {\sc Matita} proofs to {\sc HOL Light proofs}.


Instead of keeping proofs in various libraries such that that of {\sc
  HOL Light} or {\sc Matita}, and use translators, we can as well
build a single multi-system encyclopedia
such as {\sc Logipedia} and use them directly when developing a
proof in one system or the other.

\paragraph{Empirical results}

As a proof of concept, we have translated to D[Mat] the arithmetic
library of {\sc Matita} \cite{DBLP:phd/hal/Assaf15}
up to Fermat's little theorem
(around 300 lemmas). We have then translated it to D[HOLL] and
exported it to {\sc HOL Light}, {\sc Isabelle/HOL}, {\sc Coq}, {\sc
  Lean}, {\sc PVS}, and, of course, {\sc Matita}.

The proofs are available at {\tt http://logipedia.science}.

\bibliography{logipedia}

\end{document}